\documentclass[twocolumn,superscriptaddress,aps,prb]{revtex4-2}
\pdfoutput=1

\usepackage[utf8]{inputenc}
\usepackage[T1]{fontenc}
\usepackage{textgreek}
\usepackage{newtxtext,newtxmath} 

\usepackage{graphicx}
\usepackage{hyperref}
\usepackage{bm}

\begin{document}
	
\title{Anisotropic zero-resistance onset in organic superconductors}
\author{Vladislav~D.~Kochev}
\affiliation{National University of Science and Technology MISiS, 119049, Moscow, Russia}
\author{Kaushal~K.~Kesharpu}
\affiliation{National University of Science and Technology MISiS, 119049, Moscow, Russia}
\author{Pavel~D.~Grigoriev}
\email{grigorev@itp.ac.ru}
\affiliation{L.~D.~Landau Institute for Theoretical Physics, 142432, Chernogolovka, Russia} 
\affiliation{National University of Science and Technology MISiS, 119049, Moscow, Russia}
\affiliation{P.~N.~Lebedev Physical Institute, RAS, 119991, Moscow, Russia}

\begin{abstract}
We study the coexistence of superconductivity (SC) and density-wave state
and reconcile various puzzling experimental data in organic superconductors
(TMTSF)$_{2}$PF$_{6}$ and (TMTSF)$_{2}$ClO$_{4}$. The anisotropic resistance
drop above $T_c$ is qualitatively described by nascent
isolated SC islands within a bulk analytical model. However, the observed 
anisotropic SC onset is explained only when the finite size and flat needle 
shape of samples is considered. Our results pave a way to estimate 
the volume fraction and the typical size of SC islands in far from the
sample surface, and apply to many inhomogeneous superconductors,
including high-$T_c$ cuprate or Fe-based ones.
\end{abstract}
\maketitle

\section{Introduction}\label{Intro}

The interplay between various types of electronic ordering is a subject of
extensive research in condensed matter physics. It is crucial for
understanding the electronic properties of various strongly correlated
electron systems. The coexistence of charge- or spin-density wave (CDW/SDW)
and superconductivity (SC) is very common \cite{Review1Gabovich,
ReviewGabovich2002,MonceauAdvPhys} and especially important for high-$T_c$
superconductors, both cuprate \cite{XRayNatPhys2012, XRayPRL2013,
XRayCDWPRB2017} and iron-based \cite{ReviewFePnictidesAbrahams,
ReviewFePnictides2}, for transition metal dichalcogenides \cite{CDWSCNbSe2,
NbSe2NatComm} and tetrachalcogenides \cite{MonceauAdvPhys}, for organic
superconductors \cite{Ishiguro1998, AndreiLebed2008-04-23, Hc2Pressure,
Vuletic, Kang2010, ChaikinPRL2014, LeeBrownNMRDomains, LeeTriplet,
LeeBrownNMRTriplet, Gerasimenko2014, Yonezawa2018, CDWSC}. In these
materials the density wave (DW) is suppressed by some external parameter,
such as pressure or doping. The SC transition temperature $T_{c}$ is,
usually, the highest in the coexistence region near the quantum critical
point where DW disappears. The upper critical field $H_{c2}$ is often
several times higher in the coexistence region than in a pure SC phase 
\cite{Hc2Pressure, CDWSC}, suggesting possible applications of SC/DW coexistence.

The microscopic structure of SC and DW coexistence is important for
understanding the DW influence on SC properties and SC transition
temperature $T_{c}$. The DW and SC phase separation may happen in the
momentum or coordinate space. The first scenario assumes a spatially uniform
structure, when the Fermi surface is partially gapped by DW and the
ungapped parts give SC \cite{MonceauAdvPhys, GrigorievPRB2008}. The second
scenario means that SC and DW phases are spatially separated on a
microscopic or macroscopic scale, depending on the ratio of SC domain size $d$ 
and the SC coherence length $\xi_{SC}$. An example of microscopic SC
domains with size $d < \xi_{SC}$ is the soliton DW structure, where SC
emerges in the soliton walls \cite{BrazKirovaReview, SuReview,
GrigPhysicaB2009, GG, GGPRB2007, GrigPhysicaB2009}. The SC upper critical
field $H_{c2}$ may theoretically increase several times in both coexistence scenarios 
\cite{GrigorievPRB2008, GrigPhysicaB2009}.

It is yet unknown or debated how SC and DW coexist even in the relatively
weakly correlated organic superconductors, such as (TMTSF)$_{2}$PF$_{6}$ 
\cite{LeeBrownNMRDomains, Vuletic, Kang2010, ChaikinPRL2014}, 
(TMTSF)$_{2}$ClO$_{4}$ \cite{Gerasimenko2014, Yonezawa2018} or 
$\alpha$-(BEDT-TTF)$_{2}$KHg(SCN)$_{4}$ \cite{CDWSC}. 
Among these materials the most extensive and
detailed experimental data are available for (TMTSF)$_{2}$PF$_{6}$ 
\cite{Ishiguro1998, AndreiLebed2008-04-23, Hc2Pressure, Vuletic, Kang2010,
ChaikinPRL2014, LeeBrownNMRDomains, LeeTriplet, LeeBrownNMRTriplet}. This
compound attracts special attention because superconductivity there appears
on a spin-density wave background, which violates the conservation of
electron spin and, in the case of a microscopic SDW/SC coexistence, favors 
\cite{GGPRB2007, GrigorievPRB2008} the unconventional spin-triplet 
SC. The latter is supported by the observed high in-plane
upper critical field \cite{LeeTriplet}, exceeding several times the expected
paramagnetic limit, and by the NMR Knight shift measurements \cite{LeeBrownNMRTriplet}. 
However, an indisputable experimental confirmation of
a triplet SC in (TMTSF)$_{2}$PF$_{6}$ is still missing.

\begin{figure}[tb]
\centering
\includegraphics{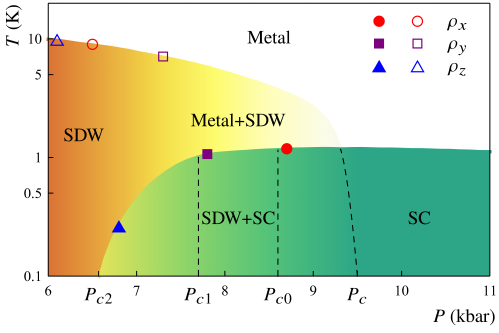}
\caption{\textbf{Pressure-temperature phase diagram} of 
(TMTSF)$_{2}$PF$_{6}$ recreated from resistivity data in Ref. \cite{Kang2010}.
Filled (blank) symbols show the transition towards SC (SDW) phase.
The intensity of green (orange) color shows the SC (SDW) volume 
fraction coexisting with the SDW (metal) phase.}
\label{FigPhDia} 
\end{figure}

At ambient pressure (TMTSF)$_{2}$PF$_{6}$ undergoes a transition from
metallic to SDW insulating state at temperature $T_{cSDW} \approx 10$ K. The
SDW transition temperature decreases with the raise of pressure 
\cite{Ishiguro1998, AndreiLebed2008-04-23, Hc2Pressure, Vuletic, Kang2010,
ChaikinPRL2014}, and SDW becomes finally suppressed at 
$P_{c} \approx 9.5$ kbar \cite{F_Pressure, Ishiguro1998,
AndreiLebed2008-04-23, Hc2Pressure, Vuletic, Kang2010, ChaikinPRL2014}, as
shown in Fig. \ref{FigPhDia}. At pressure exceeding $P_{c2} < P_{c}$
superconductivity emerges at $T < T_{c} \approx 1.1$ K. The temperature
hysteresis observed \cite{Vuletic} in the SDW/metal or SDW/SC coexistence
region in (TMTSF)$_{2}$PF$_{6}$ suggests a spatial rather than
momentum separation of the metal/SC and SDW phases. However, the origin,
size and shape of SC/metal domains in SDW phase remains unknown and debated 
\cite{Kang2010, ChaikinPRL2014}, because various observations seem to
contradict each other in a framework of any SC/SDW coexistence model. For
example, the strong increase in $H_{c2}$, both perpendicular 
\cite{Hc2Pressure, CDWSC} and parallel \cite{LeeTriplet} to conducting layers,
and the \textquotedblleft spin-triplet\textquotedblright\ SC properties 
\cite{LeeTriplet, LeeBrownNMRTriplet} suggest a microscopic SDW/SC
coexistence, e.g. the domain-wall scenario \cite{BrazKirovaReview, SuReview,
GrigPhysicaB2009, GG,GGPRB2007, GrigPhysicaB2009, Kang2010}. On the other
hand, the angular magnetoresistance oscillations (AMRO) observed in the
pressure interval of SC/SDW coexistence in (TMTSF)$_{2}$PF$_{6}$ can be
explained only by assuming a macroscopic spatial phase separation with SC
domain width $d > 1$~{\textmu}m \cite{ChaikinPRL2014}.

The most puzzling feature of SDW/SC coexistence in (TMTSF)$_{2}$PF$_{6}$, 
unexplained in any scenario, is the anisotropic SC onset 
\cite{Kang2010,ChaikinPRL2014}: with the increase in pressure at 
$P_{c2} \approx 6.7$ kbar the SC transition and the zero resistance is first observed 
only along the least-conducting
interlayer $z$-direction, then at $P_{c1} \approx 7.8$ kbar along $z$-
and $y$-directions, and only at $P_{c0} \approx 8.6$ kbar in all
directions, including the most conducting $x$-direction. This is opposite to
a weak intrinsic interlayer Josephson coupling, typical in high-$T_c$
superconductors \cite{Tinkham}. Other organic metals manifest similar
anisotropic SC onset \cite{Gerasimenko2014}. Note that the observed 
\cite{Kang2010, ChaikinPRL2014} anisotropic zero-resistance $T_{c}$ contradicts
the general rule that the percolation threshold in large heterogeneous media
must be isotropic \cite{PercolationEfros}, provided the high-conducting
inclusions are not thin filaments \cite{Kang2010} connecting opposite edges
of a sample. However, such a filament scenario cannot be substantiated
microscopically in (TMTSF)$_{2}$PF$_{6}$ and seems to be absent in the
metal/SDW coexistence region at $T > T_{c}$. Below we resolve
this paradox and reconcile relevant experimental data on SC onset in
(TMTSF)$_{2}$PF$_{6}$. The proposed model and the results obtained are 
applicable to many other superconductors and can be used to estimate 
the volume fraction and the size of SC domains.

\section{Model and calculations}\label{Calcul}

\subsection{Resistivity anisotropy above $T_c$ in large samples}\label{sec2a}

A possible clue to explain the observed SC anisotropy without invoking SC
filaments may come from a similar effects in iron selinide FeSe, where the
resistivity drop $\Delta \rho$ above $T_{c}$ was also observed to be very
anisotropic, being much greater along the least conducting interlayer
direction \cite{Sinchenko2017, Grigoriev2017}. Its superconducting origin
was confirmed by the simultaneous measurements of a diamagnetic response and
of the critical current \cite{Sinchenko2017}. This SC anisotropy was
explained within a model of a heterogeneous SC onset in the form of isolated
SC islands \cite{Sinchenko2017, Grigoriev2017}. This effect originates from
a strong conductivity anisotropy $\eta_{z} = \sigma_{zz}^{0}/\sigma
_{xx}^{0} \ll 1$ of the parent non-SC material \cite{F_Index0} and takes
place if SC islands are spheres \cite{Sinchenko2017} or even flattened
spheroids \cite{Grigoriev2017}, opposite to filaments along $z$-axis.
Isolated spherical SC islands increase conductivity in all directions
similarly, but their relative effect $\Delta \sigma_{i}/\sigma_{ii}^{0}$
for the interlayer current is $\sim 1/\eta_{z} \gg 1$ times greater than for
the in-plane current. An analytical description of this effect in fully
anisotropic compounds, i.e. with $\eta_{y} = \sigma_{yy}^{0}/\sigma_{xx}^{0} < 1$ 
and elliptic SC inclusions with main semiaxes $a_{i}$, can be
obtained using the Maxwell-Garnett approximation (MGA), valid in the limit 
of small volume fraction $\phi\ll 1$ of SC phase, or the self-consistent
approximation (SCA), describing specific spatial distributions of the 
second phase \cite{Torquato2002}. These models were derived in the bulk
limit of infinitely large samples \cite{Torquato2002}.
In MGA the resistivity $\rho_{i}=1/\sigma _{ii}$ along the axis $i \in \{x,y,z\}$
is given by \cite{Seidov2018}: 
\begin{equation}
	\rho_{i}^{MGA} = \rho_{i}^{0} \left[ \frac{A_{i}^{\ast}(1-\phi)}{%
	A_{i}^{\ast} + (1-A_{i}^{\ast})\phi }\right],  
\label{RMGA}
\end{equation}
while in SCA we obtain (details of derivation presented in Appendix \ref{appA}): 
\begin{equation}
	\rho_{i}^{SCA} = \rho_{i}^{0} \left( 1-\phi /A_{i}^{\ast} \right),  
\label{RSCA}
\end{equation}
where the diagonal components
of depolarization tensor are given by Eq. (17.25) of Ref. \onlinecite{Torquato2002}: 
\begin{equation}
A_{i}^{\ast} = \frac{1}{2}\prod\limits_{n=1}^{3}a_{n}^{\ast}
\int\limits_{0}^{\infty} \mathrm{d}t
 \left[ (t+a_{i}^{\ast 2}) \sqrt{\prod\limits_{n=1}^{3}(t+a_{n}^{\ast 2}) }\ \right]^{-1},  
\label{Ai}
\end{equation}
where $a_{i}^{\ast } = a_{i}/\sqrt{\eta_{i}}$, $\eta_{i} = \sigma_{ii}/\sigma_{xx}$.

Unfortunately, the SCA gives a qualitatively incorrect result in the limit of
strong anisotropy $\eta_{i} \ll 1$ and strong conductivity contrast 
\cite{Torquato2002}, when the conductivity of two phases differ too much, as in 
our case of SC inclusions: $\sigma^{SC}/\sigma^{0} = \infty$. This is
illustrated by our numerical calculations in 2D case shown in Fig. \ref{Fig5} (see Appendix \ref{appB}). 
From Eq. (\ref{RMGA}) one can also solve an inverse problem to
express the volume fraction $\phi$ through the conductivity with and
without SC inclusions: 
\begin{equation}
	\phi^{MGA} = \frac{A_{y} ( \sigma_{yy}-\sigma_{yy}^{0} ) }
	{\sigma_{yy}^{0}+A_{y} ( \sigma_{yy}-\sigma_{yy}^{0} ) }.  
\label{phiM}
\end{equation}

\begin{figure}[tb]
\centering



\includegraphics{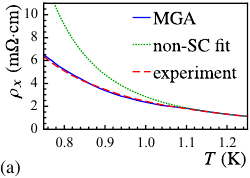}
\includegraphics{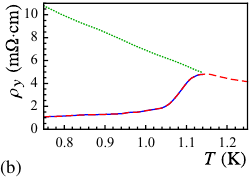}\\

\includegraphics{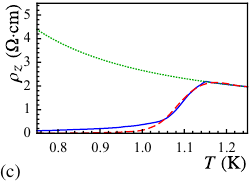}
\includegraphics{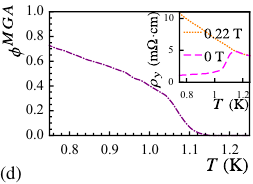}

\caption{\textbf{Temperature dependence of resistivity}
$\rho$ along (a) $x$, (b) $y$ and (c) $z$ axes.
Used experimental data for (TMTSF)$_{2}$PF$_{6}$ at $P = 8.3$ kbar were extracted
from Fig. 4a of Ref. \onlinecite{Kang2010}.
Plotted lines correspond to: calculation in MGA using Eq. (\ref{RMGA}) (solid blue); 
extrapolated resistivity before the SC onset (dotted green); experiment (dashed red).
(d) Temperature dependence of SC volume ratio calculated using Eq.
(\ref{phiM}) and experimental $\rho_{y}$ in magnetic field (inset) at $B = 0.22$ T
and $B = 0$ T.}

\label{Fig2}
\end{figure}

We apply Eqs. (\ref{RMGA})--(\ref{phiM}) to fit the observed resistivity
anisotropy $\rho_{i}(T)$ in (TMTSF)$_{2}$PF$_{6}$ \cite{Kang2010,
ChaikinPRL2014} at $T>T_{c}$ (see Fig. \ref{Fig2}). The required temperature
dependence $\phi (T)$ is extracted using Eq. (\ref{phiM}) from the
resistivity data \cite{Kang2010} without and with magnetic field destroying
SC (see Fig. \ref{Fig2}d). From Fig. \ref{Fig2} one sees that the observed
very anisotropic temperature dependence of resistivity $\rho_{i}(T)$ is
qualitatively described by isolated SC islands within MGA. The effect 
of SC inclusions on resistivity in MGA is clearly seen from the difference 
between the solid blue and dotted green curves in Fig. \ref{Fig2}, showing 
$\rho_{i}(T)$ with and without SC islands. However, the
MGA cannot explain the anisotropic zero-resistance onset observed in (TMTSF)$_{2}$PF$_{6}$ 
\cite{Kang2010, ChaikinPRL2014} and (TMTSF)$_{2}$ClO$_{4}$ \cite{Gerasimenko2014}, 
i.e. the anisotropy of SC transition temperature $T_{c}$
where the observed resistivity drops by several orders of magnitude.
Moreover, such a $T_{c}$ anisotropy seems to contradict the percolation theory 
\cite{PercolationEfros}.

\subsection{Finite-size effects and zero resistance onset}\label{sec2b}

To resolve this puzzle we note that the percolation threshold is isotropic
only in infinite heterogeneous media \cite{PercolationEfros}, i.e. when
the sample dimensions are much larger than the size $d$ of SC islands. Usually, the
single crystals of organic metals are flat whiskers elongated in the most conducting $x$-direction with a tiny 
thickness along the interlayer $z$-axis. The (TMTSF)$_{2}$PF$_{6}$
samples in the experiments of Refs. \onlinecite{Vuletic, Kang2010} were $3 \times 0.2 \times 0.1$ mm$^{3}$. 
The typical dimensions of (TMTSF)$_{2}$ClO$_{4}$
single crystals are similar: $3 \times 0.1 \times 0.03$ mm$^{3}$ in Ref. \onlinecite{Gerasimenko2014}, 
or $2.4 \times 0.7 \times 0.1$ mm$^{3}$ in Ref. \onlinecite{Yonezawa2018}. 
The observation of AMRO and FISDW in (TMTSF)$_{2}$PF$_{6}$
at field $B \approx 2$ T restricts the minimal size $d_{\min}$ of SC islands
to $d_{\min} > 1$ {\textmu}m \cite{ChaikinPRL2014}. On the other hand, the
observed \cite{Hc2Pressure, CDWSC} increase in $H_{c2}$ restricts the
maximal SC size to $d_{\max} < \lambda$, where the
penetration depth $\lambda (T = 0.19$ K$)\approx 40$~{\textmu}m in 
(TMTSF)$_{2}$ClO$_{4}$ \cite{lambdaClO4}, and a close $\lambda$ is expected 
in (TMTSF)$_{2}$PF$_{6}$ \cite{lambdaOM}. Similar $H_{c2}$ enhancement and AMRO 
were also observed in (TMTSF)$_{2}$ClO$_{4}$ \cite{Gerasimenko2013,Gerasimenko2014}.
These experimental data suggest that the typical size $d$ of SC islands in (TMTSF)$_{2}$PF$_{6}$
and (TMTSF)$_{2}$ClO$_{4}$ gets into the interval $1$~{\textmu}m $< d \lesssim 40$~{\textmu}m, 
being comparable to the sample thickness $L_{z} \sim 100$ {\textmu}m. 
Thus we need to analyze the effect of finite sample size.

\begin{figure}[tb]
\centering



\includegraphics{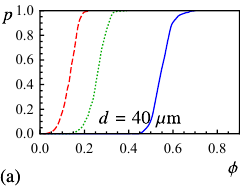}
\includegraphics{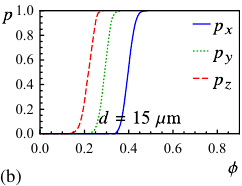}\\

\includegraphics{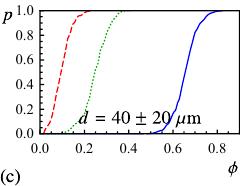}
\includegraphics{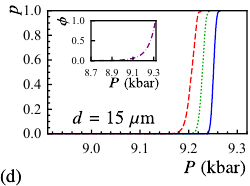}

\caption{\textbf{Percolation probability} $p$ along $x$ (solid blue), 
$y$ (dotted green) and $z$ (dashed red) axes as a function of SC volume fraction $\phi$. 
Spherical SC inclusions have diameter (a) $d = 40$ {\textmu}m, (b) $d = 15$ {\textmu}m, and 
(c) $d = 40 \pm 20$ {\textmu}m with standard deviation of $20$ {\textmu}m.
(d) Dependence of $p$ on pressure $P$ along main axes for spherical SC inclusions of 
$d = 15$ {\textmu}m, calculated from Fig. \ref{Fig3}b using the experimental data 
$\phi(P)$ (inset) extracted from Tab. 1 of Ref. \onlinecite{Vuletic}.}

\label{Fig3}
\end{figure}

For this end we calculated percolation thresholds $\phi^{c}$ numerically for randomly 
distributed spherical SC inclusions of various diameter $d$ in a sample of dimensions 
$3 \times 0.2\times 0.1$ mm$^{3}$, as in the experiment \cite{Vuletic, Kang2010}. 
For $d > 10$ {\textmu}m $\phi^{c}$ strongly depends on the distribution pattern of SC islands, 
hence, the percolation probabilities $p(\phi^{c})$ in Fig. \ref{Fig3} obtained by averaging 
over the large number of distribution patterns (see Appendix \ref{appC} for details of calculations). 
In Fig. \ref{Fig3} we see that $p$ is the largest along the shortest sample dimension in 
all cases. With the increase in SC volume fraction $\phi$ the SC transition, i.e. the 
supercurrent percolation, first appears along $z$, then along $y$, and only at much 
larger $\phi$ along the most conducting $x$-axis.
Since $\phi$ increases with pressure $P$ (see Fig. \ref{Fig3}d inset), it explains 
the anisotropic SC transition observed in Refs. \onlinecite{Kang2010,ChaikinPRL2014,Gerasimenko2014}. 
Notably, we do not need a questionable filamentary $z$-elongated shape of SC islands 
to describe these experiments: the effect emerges even for their opposite flattened shape. 
Thus, our scenario reconciles the 
relevant experimental facts on SC onset in (TMTSF)$_{2}$PF$_{6}$ and (TMTSF)$_{2}$ClO$_{4}$:
(i) the anisotropy of SC onset \cite{Kang2010, ChaikinPRL2014,Gerasimenko2014,Gerasimenko2013}, 
(ii) the observation of AMRO \cite{ChaikinPRL2014,Gerasimenko2013}, and (iii) the strong 
$H_{c2}$ enhancement in the DW/SC coexistence region \cite{Hc2Pressure,Gerasimenko2013,Gerasimenko2014}.

\begin{figure}[tb]
\centering

\includegraphics{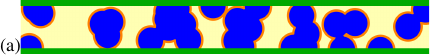}\\

\includegraphics{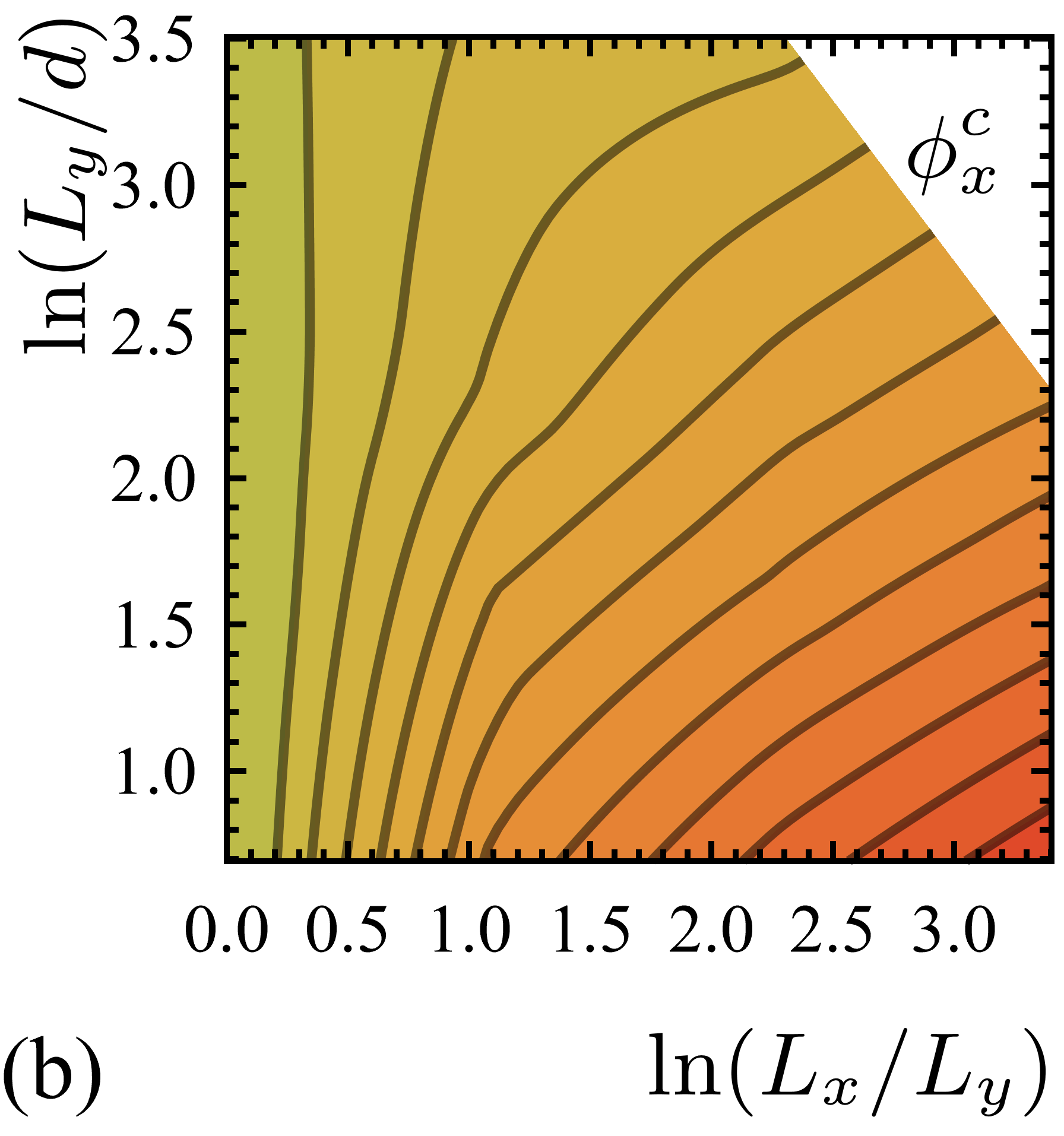}
\includegraphics{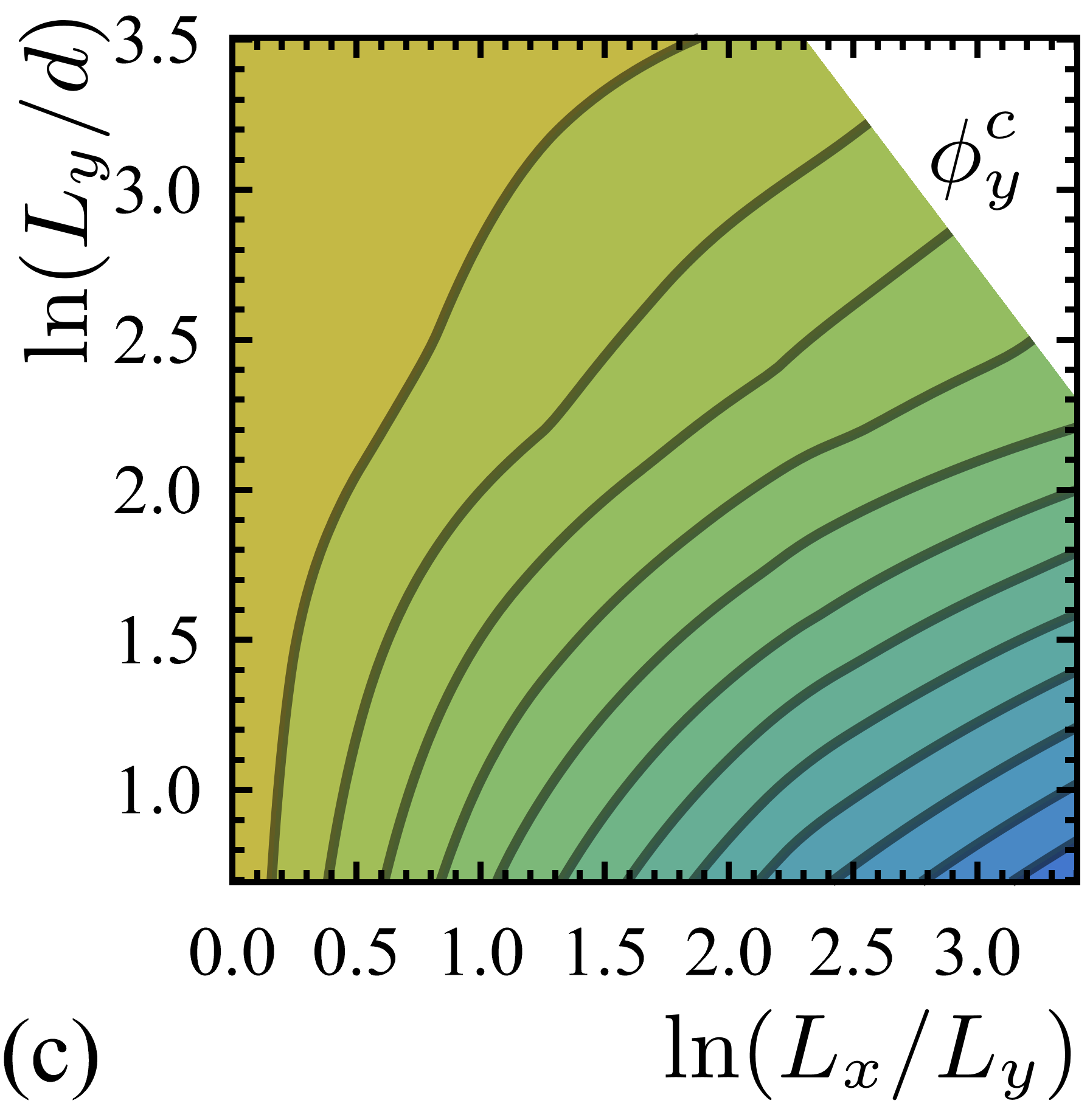}
\includegraphics{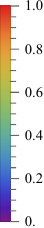}

\caption{\textbf{Dependence of percolation threshold} $\phi^{c}$ on sample shape and size in 2D.
(a) Circular SC islands (blue) with diameter $d = 0.3$ are randomly distributed inside a rectangular 
sample (yellow) of dimensions $10 \times 1$, forming SC channels between contact electrodes (green).
It shows that the percolation along sample thickness is much easier than along sample length.
(b,c) $\phi^{c}_x$, $\phi^{c}_y$ as a functions of sample length to sample thickness ratio 
$L_x / L_y$ and of sample thickness to SC island diameter ratio  $L_{y}/ d$.}

\label{Fig4}
\end{figure}

Our numerical result of anisotropic percolation threshold can be easily understood. In thin elongated
samples with $L_{x} \gg L_{z}$ the probability to find a chain of $n \approx L_{z} / d \sim 1$ 
connected SC islands, needed for percolation along the shortest edge of a sample, 
is much larger than to find a chain of length $N \approx L_{x} / d \gg 1$ for the percolation 
along the longest edge. This simple argument is illustrated in Fig. \ref{Fig4}a. 

Evidently, with the increase in sample length $L_{x}$ and thickness $L_{y}$ at other parameters $d$, $L_z$, $\phi$ fixed, 
the percolation probability $p_z$ along the sample thickness grows. 
At small $p_z \ll 1$, $p_z \propto L_x \times L_y$. 
The anisotropy of SC percolation transition also depends on the ratio of SC
grain size $d$ and of the sample thickness $L_{z}$ (see Figs. \ref{Fig3}). 
This dependence is important because it allows an experimental 
study of the typical size $d$ of SC islands in various materials and far from 
the sample boundary using resistivity measurements. 

To investigate the main features of this dependence, we calculated percolation 
probabilities $p_x$ and $p_y$ as a function of diameter $d$ of SC islands in a 
2D rectangular sample of dimension $L_{x} \times L_{y}$. The results are shown 
in Figs. \ref{Fig4}b,c. 
$\phi_{i}^{c}$ is found by solving the equation $p_{i}^{c}(\phi) = 1/2$. 
We found that both $\phi_{x}^{c}$ and $\phi_{y}^{c}$ depend weakly on $L_{x}/L_{y}$
in the limit of small grains $Ly \gg d$, but strongly when the size of grains is comparable to
thickess $L_y$. Fig. \ref{Fig3} shows a similar dependence of $p_x(\phi)$ on $d$ in 
3D case.
Hence, the anisotropy of SC onset grows when the sample becomes 
thinner and longer, and when the SC grain size $d$ increases.
This shows the importance of finite-size effects for the anisotropy of SC onset.

\section{Discussion and summary}

Due to layered crystal structure of the high-$T_c$ and organic superconductors, usually, 
thin flat samples are produced during the crystal growth \cite{Karpinski_1999}.
Temperature-dependent resistivity anisotropy is observed in most high-$T_c$ superconductors, both cuprate \cite{PhysRevLett.77.4253,PhysRevB.56.R11423,PhysRevLett.60.2194,PhysRevB.67.104512,PhysRevB.87.024509,PhysRevLett.88.137005,Zverev1998,Zverev2000} 
and Fe-based \cite{PhysRevB.82.134528,PhysRevB.81.184508,Zverev2009,Sinchenko2017,Mogilyuk2019}. 
According to our model, it depends on (i) the shape of SC islands, (ii) normal-state conductivity anisotropy, 
entering Eqs. (\ref{RMGA}-\ref{phiM}), and (iii) the sample shape and size. 
The comparison of observed resistivity anisotropy of any sign with our model gives information about 
typical size and shape of SC islands deep inside the sample. 
An anisotropic resistivity drop above $T_c$ and the applicability of Eqs. (\ref{RMGA}-\ref{phiM}) 
do not require finite-size effects and persist at $d \ll L$ and arbitrary sample shape. 
However, for anisotropic zero-resistance $T_c$ the finite sample size $L \lesssim 10^3~d$ in 
any direction is important. The doping-formed inhomogeneities in cuprates are, usually, 
$\sim 10$ nm $\ll L_z \gtrsim 1$ {\textmu}m. Nevertheless, they may form much larger clusters, 
even $\gtrsim 1$ {\textmu}m as in HgBa$_2$CuO$_{4+y}$ (see Figs. 1g or 3d of Ref. \cite{InhHgBaCuO}). 
Interplay between SC and DW also leads to larger SC domains.
Tiny and flat samples, required for anisotropic zero-resistance $T_c$, can be made artificially. 
In FeSe this led to an amazing observation \cite{Sinchenko2017,Mogilyuk2019}: in $z$-direction $T_c$ increases 
from $8$ K to $12$ K as the sample thickness decreases from $300$ nm to $\sim 50$ nm, while along 
the $a-b$ plane $T_c \approx 8$ K remains unchanged. Our model explains this effect also.   

Superconductivity onsets heterogeneously in all known high-temperature
superconductors, as confirmed by numerous scanning tunneling microscopy and spectroscopy 
measurements \cite{KresinReview2006,InhBISCCO,InhBISCO2009,InhNbN,InhCaFeAs,InhHgBaCuO,InhFeSe,InhCaFeAs2018}. 
However, these and other elaborated
experimental techniques provide detailed information about the electronic structure
at the surface, which may differ from the structure deep in the bulk. The
proposed effect allows one to estimate the typical size and shape of SC islands far from
the surface by measuring the temperature dependence of resistivity along
three main axes in the samples or artificial bridges of thickness comparable to 
or 1-3 orders less than the expected size of SC grains. This knowledge is helpful for understanding
the properties and electronic structure across the phase diagram of various
high-$T_c$ superconductors.

To summarize, we provide an explanation for the puzzling experimental data in
organic superconductors, where the SC onset is highly anisotropic. 
This explanation is based on heterogeneous SC nucleation.
Above $T_c$ the resistance decreases anisotropically because the rare SC islands 
reduce resistivity most strongly along the least conducting direction, as described analytically 
in Sec. \ref{sec2a} using the effective-medium approximation, applicable to samples of any large size and 
small fraction of SC phase. 
To explain the anisotropic zero-resistance onset or $T_c$, in Sec. \ref{sec2b} we performed 
the numerical calculation of current percolation probability via SC islands in 
finite-size samples of various shape. This calculation revealed a remarkable 
anisotropy of the percolation threshold, which depends strongly on the relative size and shape of samples and SC islands.
The proposed effect allows one to estimate the typical size and shape of SC grains far from the sample boundary using resistivity measurements. 
This effect is rather general for inhomogeneous superconductors and most pronounced when the non-SC material has anisotropic resistivity, as cuprate and Fe-based superconductors, and/or when the sample shape is anisotropic.

\begin{acknowledgments}
This article is partly supported by the Ministry of Science and Higher Education of the 
Russian Federation in the framework of Increase Competitiveness Program of MISiS, by RFBR grant \# 21-52-12027, and by 
the \textquotedblleft Basis\textquotedblright\ Foundation for development of theoretical physics and mathematics. 
V.~D.~K. acknowledges the MISiS project \# K2-2020-001, and K.~K.~K. the MISiS support project 
for young research engineers and RFBR \# 19-32-90241 \& 19-31-27001. 
P.~D.~G. acknowledges the State Assignment  \# 0033-2019-0001
and RFBR grants \# 19-02-01000 \&  21-52-12043.
\end{acknowledgments}

\appendix

\section{Derivation of resistivity of heterogeneous anisotropic compound in
	the self-consistent approximation}\label{appA}

\subsection{Isotropic case with ellipsoidal inclusions}

In the self-consistent approximation (SCA), the effect of all the material
outside any inclusion is to produce a homogeneous medium whose effective
conductivity $\sigma_{ii}^{\ast}$ is the unknown to be calculated \cite{Torquato2002}. 
The diagonal components of the effective conductivity tensor 
$\sigma_{ii}^{\ast}$ along the axis $i \in \{x,y,z\}$ of a heterogeneous media
with unidirectionally aligned \emph{isotropic ellipsoidal} inclusions
in SCA can be calculated from Eqs. (18.18) and (18.19) of Ref. \onlinecite{Torquato2002}:
\begin{equation}
	\sum_{j}^{N} 
	\frac{\phi^{j} \left( \sigma^{j}-\sigma_{ii}^{\ast} \right) \sigma_{ii}^{\ast}}
	{\sigma_{ii}^{\ast}+A_{i}^{\ast} \left( \sigma^{j}-\sigma_{ii}^{\ast} \right) } = 0,  
	\label{SCAT}
\end{equation}
where $j$ numerates the phase, $\phi^{j}$ is its volume fraction, $\sigma^{j}$ 
is its conductivity, which is assumed to be \emph{isotropic}, and the
diagonal components $A_{i}^{\ast}$ of depolarization tensor for \emph{ellipsoidal}
inclusions with semiaxes $a_{i}^{\ast}$ are given by Eq. (\ref{Ai}). 
In the next subsection we generalize these results for the case of anisotropic
conductivities $ \sigma ^{j}_{ii}$ of constituent phases. For only two
different phases, $m$ and $s$, with isotropic conductivities $\sigma ^{m}$ and 
$\sigma ^{s}$, according to Eq. (\ref{SCAT}), the effective conductivity $\sigma _{ii}^{\ast }$ 
along the axis $i$ of such a heterogeneous media satisfies the equation:
\begin{equation}
	\frac{(1-\phi )\left( \sigma ^{m}-\sigma _{ii}^{\ast }\right) \sigma
		_{ii}^{\ast }}{\sigma _{ii}^{\ast }+A_{i}^{\ast }\left( \sigma ^{m}-\sigma
		_{ii}^{\ast }\right) }+\frac{\phi \left( \sigma ^{s}-\sigma _{ii}^{\ast
		}\right) \sigma _{ii}^{\ast }}{\sigma _{ii}^{\ast }+A_{i}^{\ast }\left(
		\sigma ^{s}-\sigma _{ii}^{\ast }\right) }=0, 
	\label{SCAEq}
\end{equation}
where $\phi $ is the volume fraction of phase $s$, which in our case is superconducting (SC). 
The conductivity of SC inclusions $\sigma ^{s}\to \infty $. Then from Eq. (\ref{SCAEq}) we
obtain a simple formula for the effective conductivity: 
\begin{equation}
	\sigma _{ii}^{\ast }=\frac{\sigma ^{m}A_{i}^{\ast }}{A_{i}^{\ast }-\phi }.
	\label{sigmaSCA}
\end{equation}

\subsection{Anisotropic case with ellipsoidal inclusions}

The generalization of Eq. (\ref{sigmaSCA}) to the case of anisotropic
conductivity $\sigma^{m}$ of the parent media is performed by the mapping of
the initial anisotropic problem to an isotropic one in a similar way as used in Ref. 
\onlinecite{Seidov2018} for the derivation of effective conductivity in the
Maxwell-Garnett approximation (MGA), given by Eqs. (\ref{RMGA}) and (\ref{Ai}). 
Let $\bm{J}$ and $V$ be the current density and the electric potential
respectively in the real space, and  $\sigma_{ii}^{m}$ be the conductivity
components of the parent phase. The electrostatic continuity equation in real
space is written as:
\begin{equation}
	- \nabla \cdot \bm{J} = \sum_i \frac{\partial}{\partial r_i} \left( \sigma_{ii}^{m} \frac{\partial V}{\partial r_i} \right) = 0,
	\label{eqV}
\end{equation}
where $i \in \{x,y,z\}$. After the mapping, i.e. the change of coordinates $r_{i}$ as:
\begin{equation}
	r_{i}=r_{i}^{\ast }\sqrt{\eta _{i}},~\eta _{i}=\sigma _{ii}^{m}/\sigma
	_{xx}^{m},  
	\label{map}
\end{equation}
with the simultaneous change of conductivity to $\sigma ^{m}=\sigma _{xx}^{m}$,
Eq. (\ref{eqV}) transforms to the electrostatic continuity equation  for an
isotropic media:
\begin{equation}
	- \nabla \cdot \bm{J} = \sum_i \frac{\partial}{\partial r_i^{\ast}} \left( \sigma^{m} \frac{\partial V}{\partial r_i^{\ast}} \right) = 0.
	\label{eqV1}
\end{equation}
Coordinate dependence of the electrostatic potential $V(x,y,z)$
in an inhomogeneous medium, given by solutions of the equations (\ref{eqV}) 
or (\ref{eqV1}) with proper boundary conditions, determines the effective 
conductivity of this inhomogeneous medium. Consequently, the initial problem of
conductivity in anisotropic media with some boundary conditions can be
mapped to the conductivity problem in isotropic media with new boundary
conditions, obtained from the initial boundary conditions by anisotropic 
dilatation given in Eq. (\ref{map}). These boundary conditions are
determined both by the sample boundaries and by the inclusions of second
phase. If these inclusions have ellipsoidal shape with the principal
semiaxes $a_{i}$, then after the mapping to the isotropic media these
inclusions keep an ellipsoidal shape but change the principal semiaxes to:
\begin{equation}
	a_{i}^{\ast }=a_{i}/\sqrt{\eta _{i}}.  
	\label{eq:shape}
\end{equation}
Eqs. (\ref{sigmaSCA}) and (\ref{Ai}) with semiaxes $a_{i}^{\ast}$ give the
effective conductivity in the mapped space. Making the reverse mapping to the
real space, we obtain the effective conductivity of initial heterogeneous media 
in real space in SCA:
\begin{equation}
	\sigma _{ii}=\frac{\sigma _{ii}^{m}A_{i}^{\ast }}{A_{i}^{\ast }-\phi },
	\label{sigmaSCAR}
\end{equation}%
which gives Eq. (\ref{RSCA}). 

Note that in the final formula (\ref{sigmaSCAR}) 
the effective conductivity $\sigma _{ii}$ in the real space depends on the 
parameters $A_{i}^{\ast }$ and $a_{i}^{\ast }$ in the mapped space. 
This is because the coordinate dependence of 
electrostatic potential $V\left( r_i \right) $ in the real space is obtained from 
the electrostatic potential $V^{\ast }\left( r_i^{\ast} \right) $ in the 
mapped space (with semiaxes $a_{i}^{\ast }$) via the simple substitution of 
Eq. (\ref{map}): $V\left( r_i \right) = V^{\ast }\left( r_i^{\ast} \right) $. 
The dilatation $r_i\to r_i/\sqrt{\eta_i} $ changes $\sqrt{\eta_i}$ times the electric 
field $E_i\left( r_i \right) =-\nabla_i V\left( r_i \right)$, while the electric 
current $J_i=\sigma_{ii}E_i$ changes  $1/\sqrt{\eta_i}$  times, because the local 
conductivity $\sigma_{ii}^{m}$ changes $1/\eta_i $ times. The effective conductivity 
$\sigma_{ii}$ also changes $1/\eta_i $ times: $\sigma_{ii}=J_i \bar{E}_i $, where 
the averaged (over sample size $L_i$) electric field 
$\bar{E}_i=L_i/\left[V(r_i=0)-V(r_i=L_i)\right] $ changes $1/\sqrt{\eta_i}$ 
times due to the dilatation.

\section{Comparison of the results of bulk analytical models and numerical
	calculations}\label{appB}

\begin{figure}[tb]
	\centering
	
	\includegraphics{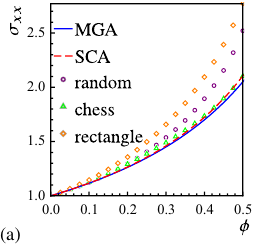}
	\includegraphics{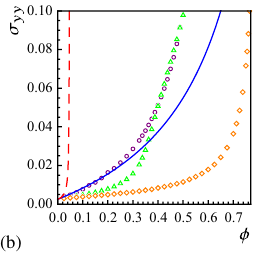}

	\caption{The conductivity of an anisotropic (square $1 \times 1$) heterogeneous media 
		with superconducting inclusions calculated using analytical models, Eq. (1) for MGA 
		and Eq. (2) or (\ref{sigmaSCAR}) for SCA, and numerically for three different 
		distributions of SC islands: random, rectangular and chess order. In SCA 
		$\sigma_{yy}(\phi)$ goes up sharply and even diverges at $\phi \sim \eta \ll 1$, 
		which drastically contradicts the numerical results.}
	\label{Fig5}
\end{figure}

In this section we compare the results, given by analytical formulas (1)-(3)
obtained in the Maxwell-Garnett (MGA) and self-consistent (SCA)
approximations, with the numerical calculations in 2D case (see
Fig. \ref{Fig5}). This allows to estimate the applicability of these
two bulk analytical models to describe real experiments on conductivity in
heterogeneous superconductors. The calculated conductivity along two axes, 
$x$ and $y$, for a square heterogeneous media of conductivity $\sigma^0_{xx} = 1$ 
and $\sigma^0_{yy} = \eta = 1 / 400$ with circular superconducting
islands as a function of their volume fraction $\phi $ is shown in Fig. \ref{Fig5}. 
For numerical calculations three different distributions of SC
islands are considered: random, rectangular and chess order. 
For rectangle order our numerical calculations give the largest 
conductivity $\sigma_{xx}(\phi)$ along the most conducting axis and the 
smallest conductivity $\sigma_{yy}(\phi)$ along the most conducting direction.
For conductivity $\sigma_{xx}(\phi)$ all approximations, both numerical 
and analytical, give similar results (see Fig. \ref{Fig5}a).
However, $\sigma_{yy}(\phi)$ in various approximations differ much, as shown in Fig. \ref{Fig5}b.
The numerical calculations of $\sigma_{yy} (\phi)$ for all three 
distributions of SC islands give rather close results, but
the analytical models MGA and SCA differ very strongly. The MGA
approximation for $\sigma_{yy} (\phi)$ is much closer to the numerical results than SCA: 
the conductivity $\sigma_{yy}$ in SCA deviates crucially and diverges at 
$\phi \sim \eta\ll 1$. This calculation illustrates the known fact \cite{Torquato2002}
that SCA, usually, gives qualitatively incorrect results in the limit of strong contrast
between the conductivities of two phases in heterogeneous media, especially in
the limit of strong anisotropy.

\section{Details of fits and calculations}\label{appC}

In plotting Fig. \ref{Fig2}d we assume that the magnetic field $B_{z} = 0.22$~T is 
strong enough to suppress superconductivity. 
In fact, such a field at $P = 8.3$ kbar reduces the SC transition temperature 
from $T_{c}(B_{z}=0) \approx 1.1$~K to $T_{c}(B_{z} = 0.22$~T$) \approx 0.3$~K. 
Hence, these data can be used to determine $\phi (T > 0.3$~K$)$. 
The magnetic field $B_{z} = 0.22$~T also leads to small metallic magnetoresistance 
$\rho_{b}(B)$, which is almost temperature independent at $T_{c} < T < 1.5$~K 
(see Fig. 4b of Ref. \onlinecite{Kang2010}). 
Therefore we take it into account in our calculation of $\phi(T)$ by 
the offset $\rho_{y}^{0}(T) = \rho_{y}(T,B_{z} = 0.22$~T$)-[\rho
_{y}(T = 1.15$~K$, B_{z} = 0.22$~T$)-\rho_{y}(T = 1.15$~K$, B_{z} = 0)]$.

The percolation probability in Figs. \ref{Fig3}, \ref{Fig4} was calculated numerically using Monte-Carlo simulation. 
For each distribution of diameters $d = \mu \pm \sigma$, which is taken Gaussian with a half-width $\sigma$,
a random state with  proper number of spherical inclusions in a box with given dimensions 
($L_x \times L_y \times L_z = 3 \times 0.2 \times 0.1$ mm$^{3}$ in 
Fig. \ref{Fig3} and various $L_x \times L_y$ in Fig. \ref{Fig4}) was generated. 
The number of SC inclusions is determined by the fixed volume fraction $\phi$ of SC phase.
Each state is associated with a graph whose vertices are SC islands. 
The vertices of the graph are connected by edges if the corresponding inclusions overlap. 
Thus, the problem of detecting the presence of percolation is reduced to finding the 
connected components of the graph, which contain the vertices 
corresponding to SC inclusions on the opposite sample edges.
For each state along each axis the percolation, i.e. the existence of a continuous path 
via intersecting inclusions, was checked, and the averaging over random realizations was made.
Depending on the parameters, from $10^4$ to $10^5$ generated realizations were enough 
to estimate the average probability of percolation in our calculations.

The conductivity of an anisotropic media (in Fig. \ref{Fig5}) was calculated 
numerically by solving the electrostatic continuity equation (\ref{eqV}) 
for the heterogeneous medium using the finite element method.  

A quantitative comparison with experiment requires the exact functions 
$\phi(P)$ and $\phi(T)$, which are known only approximately. 
Fig. 2d, based on resistivity in MGA, overestimates $\phi(T)$, because MGA gives a lower 
bound of conductivity in heterogeneous media \cite{Torquato2002}. On contrary, Fig. 3c inset, 
based on the resistivity fit above $T_{c}$ in the metal/SDW phase \cite{Vuletic}, 
underestimates $\phi(P)$, because the volume fraction of SC phase at $T < T_{c}$ 
should be larger than the volume fraction of metal phase at $T_{c} < T \ll T_{cSDW}$ 
for two reasons: (i) superconducting phase has lower energy than metallic 
phase, and (ii) the SC proximity effect increases the effective SC volume fraction. 

\bibliographystyle{apsrev4-2}
%

\end{document}